\documentclass[pra,twocolumn,showpacs,superscriptaddress]{revtex4-1}
\usepackage{amsmath,amscd,amssymb,color}
\usepackage{graphicx,amsfonts,epsf}
\usepackage{epstopdf}
\usepackage{float}
\usepackage{hyperref}
\usepackage{enumerate}
\usepackage{array}

\usepackage{xcolor}
\usepackage{breakurl}
\begin{document}
\title{Experimental investigation of a quantum Otto heat
engine with shortcuts to adiabaticity implemented using counter-adiabatic
driving}
\author{Krishna Shende}
\email{ph19032@iisermohali.ac.in}
\affiliation{Department of Physical Sciences, Indian
Institute of Science Education \& 
Research Mohali, Sector 81 SAS Nagar, 
Manauli PO 140306 Punjab India.}
\author{Matreyee Kandpal}
\email{ph23022@iisermohali.ac.in}
\affiliation{Department of Physical Sciences, Indian
Institute of Science Education \& 
Research Mohali, Sector 81 SAS Nagar, 
Manauli PO 140306 Punjab India.}
\author{Arvind}
\email{arvind@iisermohali.ac.in}
\affiliation{Department of Physical Sciences, Indian
Institute of Science Education \& 
Research Mohali, Sector 81 SAS Nagar, 
Manauli PO 140306 Punjab India.}
\author{Kavita Dorai}
\email{kavita@iisermohali.ac.in}
\affiliation{Department of Physical Sciences, Indian
Institute of Science Education \& 
Research Mohali, Sector 81 SAS Nagar, 
Manauli PO 140306 Punjab India.}
\begin{abstract}
The finite time operation of a quantum Otto heat engine leads to a trade-off
between efficiency and output power, which is due to the deviation of
the system from the adiabatic path. This trade-off caveat can be
bypassed by using  the shortcut-to-adiabaticity protocol. We
experimentally implemented a quantum Otto heat engine using spin-1/2
nuclei on a nuclear magnetic resonance (NMR) quantum processor.  We
investigated its performance using the shortcut-to-adiabaticity
technique via counter-adiabatic driving with the inclusion of the cost
to perform the shortcut.  We use two different metrics that incorporate
the cost of shortcut-to-adiabaticity to define engine efficiency and
experimentally analyze which one is more appropriate for the NMR
platform.  We found a significant improvement in the performance of the
quantum Otto heat engine driven by shortcut-to-adiabaticity, as
compared to the non-adiabatic heat engine.
\end{abstract} 
\maketitle 
\section{Introduction}
\label{intro}
One of the fundamental results of thermodynamics is that the efficiency of any
heat engine with hot and cold reservoirs (at temperature $T_h$ and $T_c$,
respectively), is limited by the Carnot efficiency( $\eta_C=1-T_c/T_h$). Since
physical heat engines have a finite operation cycle time, they are not in an
exact equilibrium state during operation, and hence do not satisfy the
quasi-static process condition.  Consequently, their efficiency is always lower
than the maximum achievable Carnot efficiency.  Quantum thermodynamics, which
deals with the thermodynamics of quantum systems, has made rapid progress in
recent years and fluctuation theorems and thermodynamic uncertainty relations
have been discovered~\cite{Deffner_book,Sai_2016}. A heat engine operating on
the quantum scale is called a quantum heat engine (QHE), and experimental
implementations of QHEs have been reported in
NMR~\cite{Peterson_2019,Assis_2019,Measure_engine}, superconducting
qubits\cite{PRX_IBM}, NV centers~\cite{NV_engine_2019} and ion
traps~\cite{Trap1,Trap2}.  A QHE has been proposed with an Ising spin chain as
the working material~\cite{Piccitto_2022,Victor_2022}, the effect of coherence
in QHEs has been studied\cite{Serra_Cohe}, and an engine driven by quantum
coherence has been proposed~\cite{Aimet_2023}.

The working of the QHE is based on the assumption that ideal adiabatic
transformations are quasi-static, which implies that the QHE cycle
time is infinitely long, to achieve an ideal process. In real life,
QHEs are implemented in a finite time, which leads to
irreversibility, which can be quantified as irreversible work and inner friction
in the thermodynamic process~\cite{Plastina_2014}. The
non-commutativity of the driving Hamiltonian at different times induces
transitions among the instantaneous energy eigenstates, which is classified as
quantum friction~\cite{Kosloff_2006,Thomas_2014,Kosloff_2022}.  The finite time
operation of a practical QHE and extra coherence build up due to
non-commutativity of Hamiltonian leads to a trade-off between power and
efficiency, such that an increase in efficiency leads to a decrease in power and
vice versa. 

A long standing hurdle in experimental implementations of QHEs is to design an
efficient QHE which can deliver more output in a finite operation time.  A new
quantum control protocol, called short to adiabaticity(STA) was proposed to
overcome this difficulty~\cite{Chen_2010,Chen_STA,STA_review,Lutz_2017}.  In an
STA protocol, a quantum adiabatic process is mimicked in a finite amount of
time such that there is no transition between the energy eigen states during
the transformation and thus the system follows an adiabatic path.  Various
approaches to implement STA have been  proposed, including via
counter-adiabatic driving, where an extra term is added to the original
Hamiltonian which keeps the system on an adiabatic path and suppresses the
non-adiabatic
transitions~\cite{Demirplak2003,Demirplak2005,Berry_2009,Takahashi_2013}.  STA
has been successfully implemented experimentally on various quantum processing
platforms such as superconducting qubits\cite{Funo_2019}, Bose-Einstein
condensate\cite{Bason2012}, NV centers\cite{Zhang_2013}, cold
atoms\cite{Du2016}, trapped ions\cite{An2016} and NMR\cite{Suresh_2023}.

In this work, we use counter-adiabatic driving in a spin-1/2 Landau-Zener(LZ)
model of a quantum Otto heat engine to implement STA, and explore its
performance on an NMR quantum processor.  We experimentally implement the
engine cycles using two qubits for two different sets of reservoir
temperatures.  We also compared the figures of merit of the quantum Otto heat
engine, implemented with and without STA, in order to investigate the dynamics
of efficiency and power. Engine efficiency was calculated by taking into
account STA driving costs.  We computed engine efficiency using two different
definitions that incorporate the STA cost, and experimentally determine the
definition which is useful for the NMR platform.  Our experimental results
conclude that the quantum Otto heat engine performs better with STA driving as
compared to the nonadiabatic quantum heat engine.

The rest of this paper is organized as follows:~Section~\ref{theory} contains a
brief description of the requisite theoretical background, with
Sections~\ref{2a}-\ref{2c} describing the quantum Otto heat engine, the basics
of counter-adiabatic driving, and the engine model, respectively.
Section~\ref{sec3} describes the experimental implementation of the heat
engine.  Section~\ref{results} contains a discussion of the main results, while
Section~\ref{concl} offers some concluding remarks.
\section{Theoretical Background}
\label{theory}
\subsection{Quantum Otto heat engine}
\label{2a}
\begin{figure}[ht]
\includegraphics[scale=1]{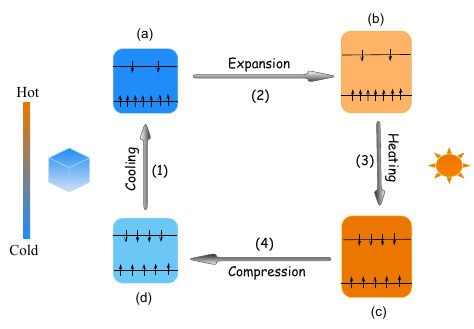}
\caption{Schematic representation of a quantum Otto heat
engine cycle. (1) and (3) are cooling and heating strokes
which are implemented by establishing contact between a hot and a cold
bath respectively, which results in a change in
the populations of the working system. (2) and (4) are expansion and
compression strokes which are realized by changing the energy
gap of the working system. } 
\label{heat engine}
\end{figure}

The quantum Otto heat engine cycle is a four stroke cycle
operating between a hot and a cold reservoir.
Two strokes are isochoric and two strokes are
adiabatic.  The isochoric processes are carried out by
thermalizing the system with a hot or a cold
reservoir, as needed. The adiabatic processes comprise
of an expansion and a compression and
are carried out by changing the time-dependent
parameter($\lambda(t)$) of the system Hamiltonian
$H_0(\lambda(t))$. The
schematic diagram of quantum Otto heat engine
cycle is shown in Figure.\ref{heat engine}. 

\vspace*{6pt}
\noindent
{(i)}\textit{Cooling}: At first, the working medium is brought
in contact with a cold bath at inverse temperature $\beta_C$
such that it is equilibrated to a state $\rho_a=e^{-\beta_C
H(\lambda_1)}/{Z_1}$, where $H(\lambda_1)$ is the system
Hamiltonian and $Z_1$ is the partition function. The cold
inverse temperature has the form $\beta_C=1/k_BT_C$, where
$k_B$ is the Boltzmann constant and $T_C$ is the cold
bath temperature.

\vspace*{6pt}
\noindent
(ii)\textit{Expansion}: In this step, the system is separated
from bath and the Hamiltonian parameter is changed from
$H(\lambda_1)$ at time t=0 to $H(\lambda_2)$ till time
t=$\tau$ using a unitary transformation. This increases the
energy gap between the working medium adiabatically and
is assumed to be ideal, such that no unwanted transitions 
occur during evolution.

\vspace*{6pt}
\noindent
(iii)\textit{Heating}: Again, contact between the hot
bath at inverse temperature $\beta_H$ and the working medium is
established, such that the system thermalizes to a state
$\rho_c=e^{-\beta_H H(\lambda_2)}/{Z_2}$, where $Z_2$ is the
partition function.

\vspace*{6pt}
\noindent
(iv)\textit{Compression}: In this step, the time
dependent Hamiltonian parameter $\lambda(t)$ of the working
medium is changed from $\lambda_2$ to $\lambda_1$ under a
unitary evolution.

The average work performed during unitary adiabatic driving
i.e. expansion or compression is given by~\cite{Deffner_book,Sai_2016}:
\begin{equation}
\langle W \rangle = Tr[H_f\rho_f]-Tr[H_i\rho_i]
\end{equation}
where $H_i$ and $\rho_i$ are the initial Hamiltonian and density
matrix immediately after the thermalization step, and  $H_f$ and
$\rho_f$ are the final Hamiltonian and the density matrix after
the expansion or compression process.

The average heat exchange between the reservoir and the
system is given as:
\begin{equation}
\langle Q \rangle =
Tr[H^\prime\rho^{\prime}_f]-Tr[H^\prime\rho^{\prime}_i]
\end{equation}
where $H^\prime$ is the system Hamiltonian during
the thermalization process, which remains constant;
$\rho^{\prime}_i$ and $\rho^{\prime}_f$ are the density matrix
before and after the thermalization process, respectively.
\subsection{Counter-adiabatic driving}
\label{2b}
When a system evolves under a time dependent Hamiltonian $H_0(t)$, the
Hamiltonian must vary slowly enough so that the system evolves adiabatically
and  remains in the instantaneous eigen state of the Hamiltonian. A Hamiltonian
not obeying the adiabaticity condition will induce unwanted transitions during
the evolution.
These unwanted transitions can be avoided by using an STA
technique such as counter-adiabatic
Driving(CD)~\cite{Berry_2009,Takahashi_2013}. In
this method, an additional term ($H_{CD}(t)$) is
added to the original Hamiltonian($H_{0}(t)$): 
\begin{equation}
H(t)=H_{0}(t)+H_{CD}(t)
\end{equation} 
The CD scheme ensures that the system remains in the
instantaneous eigenstates of the original Hamiltonian $H_0(t)$
when it is driven using the new effective Hamiltonian $H(t)$.
Hence, the system follows an adiabatic path.
The exact form of
$H_{CD}(t)$ is given by~\cite{Berry_2009}:
\begin{equation}
H_{CD}(t)=i\hbar\sum_{n}(|\partial_tn\rangle\langle
n|-|n\rangle\langle\partial_t n|n\rangle\langle n|)
\end{equation}
where $|n\rangle = |n(t)\rangle$ is the n$^{th}$ eigenstate
of the original Hamiltonian $H_0(t)$, at time $t$.
\subsection{Engine model}
\label{2c}
For a spin-1/2 system placed in an arbitrary magnetic field
and assuming a Landau-Zener type model, the original
Hamiltonian is given by~\cite{Zener}:
\begin{equation}
H_{0}(t)=b_x\sigma_x+b_z(t)\sigma_z
\end{equation}
where $b_x$ and $b_z(t)$ are external magnetic fields, $b_x$
is the minimum splitting frequency between energy levels,
$b_z(t)$ is the time-dependent external field and
$\sigma_{x,z}$ are the Pauli matrices. The CD
Hamiltonian assumes the form: 
\begin{equation}
H_{CD}(t)=-\dfrac{b_x
\dot{b_z}}{2[b_{x}^{2}+b_{z}^{2}]}\sigma_y=b_{CD}(t)\sigma_y
\label{H_cd}
\end{equation}
The form of $b_j(t)$ is given by~\cite{Cakmak_2019}:
\begin{equation}
b_j(t)=C_j+D_j\dfrac{t^2}{\tau^2}\left(\frac{1}{2}-\dfrac{t}{3\tau}\right)
\label{externalfield}
\end{equation}
where $C_j$ and $D_j$ are arbitrary constants that
determine the initial and final values of the external
driving frequency, \textit{j} is either x or y, $\tau$ is
the total driving time and $t$ $\in$ [0,$\tau$].
The CD Hamiltonian $H_{CD}$ must vanish at the beginning and
at the end of the driving time, which imposes the conditions
$b_{CD}(t=0,\tau)$=0 and $\dot{b}_{CD}(t=0,\tau)$=0. The
form of Eq.\ref{externalfield} already takes care of the
above boundary conditions, and hence can be used for STA
driving.

The performance of a heat engine is quantified by its working efficiency and
output power. Engine efficiency is quantified by the net output work performed
by the working medium divided by the heat transferred from the hot bath, giving
the fraction of heat energy converted into work.  Output power is the rate at
which output energy is produced.  Ideally, for an adiabatic cycle, the
efficiency and power are given by:
\begin{equation}
\eta_A=-\dfrac{\langle W_2 \rangle+\langle
W_4\rangle}{\langle Q_3 \rangle} \hspace{0.5cm} \&
\hspace{0.5cm} P_A=-\dfrac{\langle W_2 \rangle+\langle
W_4\rangle}{\tau_{cycle}}
\end{equation}
where $\eta_A$ and $P_A$ are the adiabatic
efficiency and power respectively, $\langle W_2 \rangle$ and
$\langle W_4 \rangle$ denote average work performed during
the expansion and compression steps, $\langle Q_3 \rangle$ is
the average heat absorbed during the isochoric heating step, and
$\tau_{cycle}$ is the cycle duration.

For an STA heat engine, an extra term has 
been added to the original Hamiltonian
to enhance its overall working dynamics. Hence,
this extra input energy needs to be considered in the performance
of the STA heat engine.  

The first definition of the efficiency of the STA
engine is given by~\cite{Cakmak_2019,Lutz_2017,Obinna_2019}:
\begin{equation}
\eta^1_{STA}=-\dfrac{\langle W_2 \rangle^{STA}+\langle W_4
\rangle^{STA}}{\langle Q_3 \rangle + \langle \dot{H}_2
\rangle^{STA}_{\tau} + \langle \dot{H}_4 \rangle^{STA}_{\tau}}
\label{first_STA_def}
\end{equation}
Similarly, the second definition for the efficiency of the STA engine has been
proposed in Ref.~\cite{Gilchrist_2019} as:
\begin{equation}
\eta^2_{STA}=-\dfrac{\langle W_2 \rangle^{STA}+\langle W_4 \rangle^{STA} +
\langle \dot{H}_2 \rangle^{STA}_{\tau} + \langle \dot{H}_4
\rangle^{STA}_{\tau}}{\langle Q_3\rangle }  
\label{second_STA_def}
\end{equation}
and the power is given by:
\begin{equation}
P_{STA}=-\dfrac{\langle W_2 \rangle^{STA}+\langle W_4
\rangle^{STA} - \langle \dot{H}_2 \rangle^{STA}_{\tau} -
\langle \dot{H}_4 \rangle^{STA}_{\tau}}{\tau_{cycle}}
\label{STA_power}
\end{equation}
where STA denotes quantities that define the STA heat engine
and the cost of the STA implementation during an adiabatic process
is given by~\cite{Cakmak_2019}:
\begin{equation}
\langle \dot{H}_{i} \rangle^{STA}=\int_{0}^{\tau}\langle
\dot{H}_{CD}(t) \rangle^{STA} dt
\label{STA_cost}
\end{equation}
which is the time average of the energy spent to implement STA
($i=2,4$), and H$_{CD}$ is the form of the CD Hamiltonian during
expansion or compression. The above equations
define the energy required to maintain an adiabatic path during
expansion and compression.
\section{Experimental implementation}
\label{sec3}
\subsection{Experimental details}
\label{3a}
\begin{figure}[ht]
\includegraphics[scale=1]{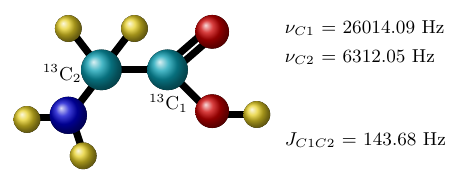}
\caption{Molecular  structure  of $^{13}$C$_2$-labeled glycine  with the
$^{13}$C$_1$ and $^{13}$C$_2$ nuclei encoded as the first and second
qubits, respectively. The scalar coupling strength and the offset
rotation frequencies, are listed alongside.}
\label{glycine}
\end{figure}

In our experiments, we have used $^{13}$C$_2$-labeled
Glycine where the $^{13}$C$_1$ and $^{13}$C$_2$ spins are
encoded as the first and second qubits, respectively. The  molecular  structure
and system  parameters  including  the  chemical  shifts
$\nu_i$ and the scalar coupling constant $J$ are  given  in
Fig.~\ref{glycine}.
The rotating frame Hamiltonian for two spin-1/2 nuclei is
given by~\cite{OLIVEIRA}:
\begin{equation}
\mathcal{H}=-\hbar\sum\limits_{i=1}^2 { \omega _i } I_z^i +
\hbar\sum\limits_{i<j=1}^2  J_{ij} I_z^i I_z^j
\end{equation}
where $\omega_i$ the offset angular frequency of the \textit{i}$^{th}$ nucleus,
$I_z^i$ represents the z-component of the spin angular momentum of the
\textit{i}$^{{\rm th}}$ nucleus and $J_{ij}$ is the scalar coupling between the
\textit{i}$^{{\rm th}}$ and the \textit{j}$^{{\rm th}}$ nuclei.  All the
experiments  were performed  at  room  temperature on a Bruker Avance III
600-MHz FT-NMR spectrometer equipped with a QXI probe.  The measured T$_1$ and
T$_2$ relaxation times the for $^{13}$C$_1$ and $^{13}$C$_2$ spins are 28.57~s
and 3.28~s and are 1.84~s and 1.25~s, respectively.

\begin{figure*}[ht]
\includegraphics[scale=1]{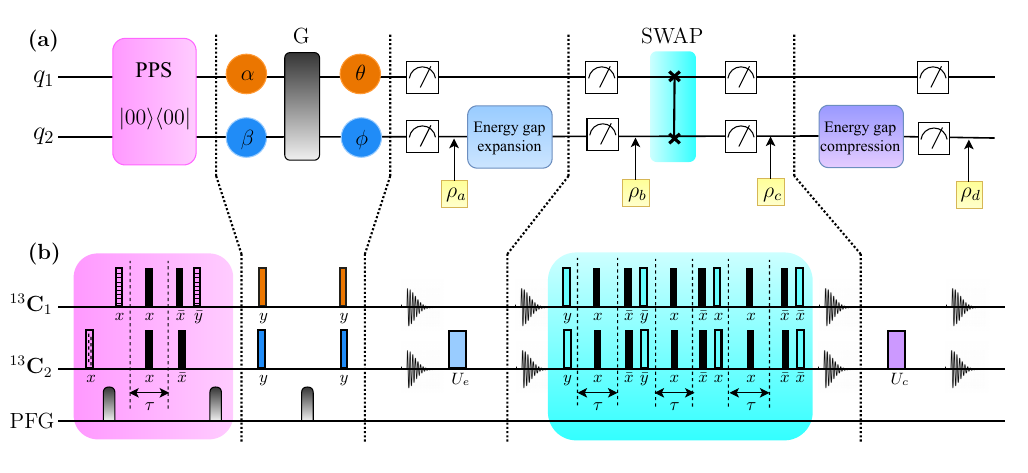}
\caption{(a)Schematic circuit diagram to realize QOHE using
a two-qubit system in NMR. (b) NMR pulse sequence to
implement the Otto heat engine. Bars containing
dashed vertical lines, bars with horizontal lines, unfilled
and filled bars represent ($\frac{\pi}{3}$),
($\frac{\pi}{4}$), ($\frac{\pi}{2}$) and ($\pi$) rotation
angles respectively, with their corresponding rotation axis given below
each bar. Orange and blue bars represent pulses used to
initialize hot and cold bath temperatures, respectively. Cyan
and violet bars depict external driving
expansion(U$_e$) and compression(U$_c$) unitary pulses,
respectively. The line depicting application of
a pulsed field gradient (PFG) marks the times at which
a gradient is employed to destroy coherence. The time delay
$\tau$ is set equal to $\frac{1}{2J_{C_1 C_2}}$.   }
\label{ckt_diag}
\end{figure*}
 
The complete circuit diagram and pulse sequence to perform
the study of non-equilibrium dynamics is shown in
Fig.\ref{ckt_diag}. The spatial averaging
technique\cite{CORY1998} has been used to initialize the system in a
pseudopure state $|00\rangle \langle 00|$ with the corresponding density
operator given by:
\begin{equation}
\rho_{00}=\dfrac{1-\epsilon}{4}I_4+\epsilon|00\rangle \langle 00|
\end{equation}
where $I_4$ is the 4 $\times$ 4 identity operator and $\epsilon$ is
proportional to the spin polarization( $\approx$ 10$^{-5}$ at room
temperature). The circuit and pulse sequence used to achieve this are shown in
the pink shaded portion of Fig.~\ref{ckt_diag}.  We have used the Gradient
Ascent Pulse Engineering (GRAPE) technique\cite{Khaneja_2005} for the
optimization  of all  the  radiofrequency(RF) pulses  used  to  construct  the
pseudopure state.  The  GRAPE optimized RF pulses are robust against RF
inhomogeneity,  with  an  average  fidelity  of 0.999. All the $\frac{\pi}{2}$
pulses used to create the pseudopure state have a duration of 150 $\mu$s with a
maximum power level of 179.47~W for the QXI probe. 
 
We  performed  full quantum state tomography to reconstruct the experimental
matrix. The closeness between the
theoretical and experimental states is quantified using the fidelity
measure~\cite{Zhang_f,WANG_2008}: 
\begin{equation} 
F=\frac{|Tr(\rho_e
\rho_t^{\dagger})|}{\sqrt{Tr(\rho_e\rho_e^{\dagger})
Tr(\rho_t\rho_t^{\dagger})}}
\end{equation} 
where $\rho_t$ and $\rho_e$ are the theoretically predicted 
and experimentally measured
density matrices, respectively.
 
\begin{figure*}[ht] 
\includegraphics[scale=1]{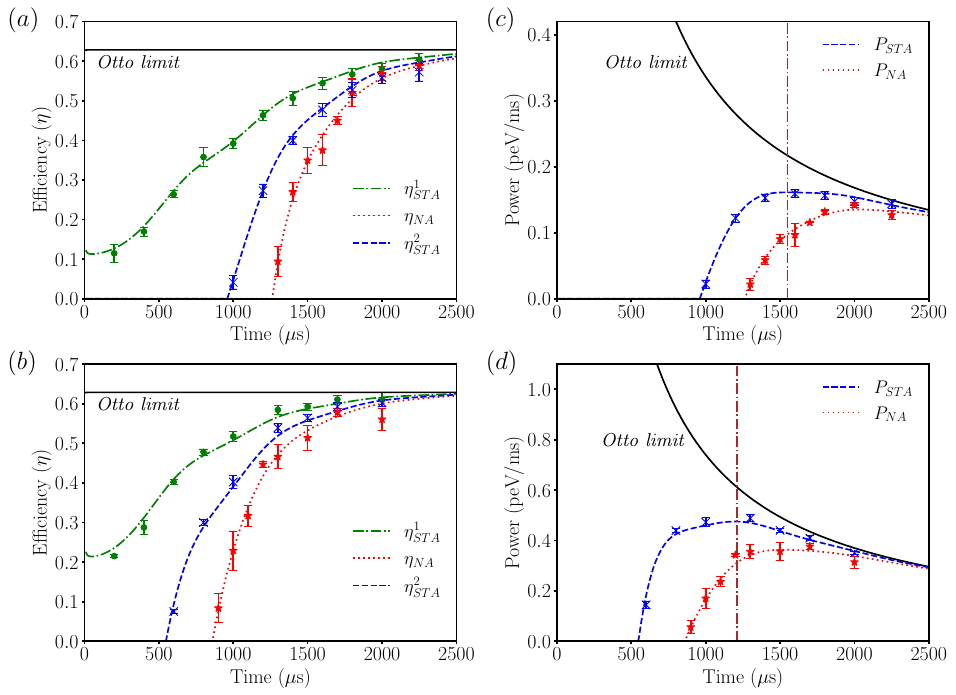}
\caption{Quantum Otto heat engine dynamics: Efficiency((a)-(b)) and
output power((c)-(d)) as a function of driving time($\tau$). Solid
black lines mark the theoretical prediction, while dashed blue,
dot-dashed green and dotted red curves are from the simulations.
Points marked with blue crosses, green circles and red stars (along with
error bars) represent experimental data.  The cold source spin
temperature is set to $k_BT_C$=11.94 peV.  Plots in the upper and lower
panels correspond to hot spin temperatures of $k_BT_H^1$=40.54 peV and
$k_BT_H^2$=53.11 peV, respectively.  The driving time at which maximum power is
obtained is marked by a dot-dashed vertical line.  } 
\label{effi_plots}
\end{figure*}

\begin{figure}[ht]
\includegraphics[scale=1]{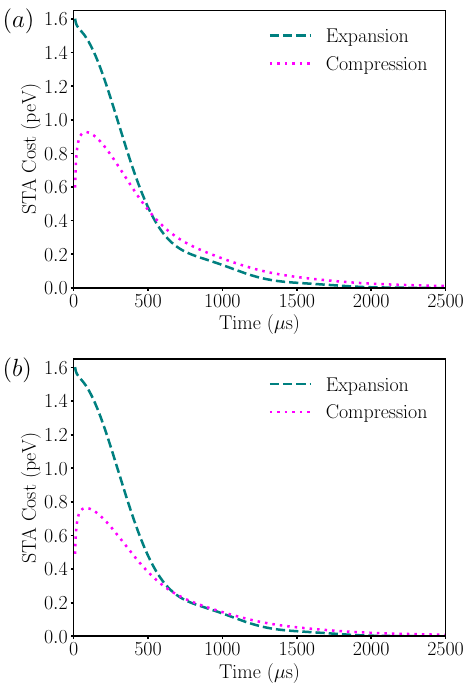} 
\caption{Quantum Otto Engine STA cost dynamics: The theoretically
calculated STA cost for the expansion and compression processes is
depicted using dashed teal and dotted magenta lines, respectively, as a
function of the driving time (\(\tau\)). The cold source spin
temperature is fixed at \(k_B T_C = 1.9\) peV. The upper and lower
plots correspond to hot spin temperatures of \(k_B T_{H1} = 6.45\)
peV and \(k_B T_{H2} = 8.45\) peV, respectively.}
\label{STA_cost_plots}
\end{figure}
\subsection{Experimental implementation of the Otto engine}
\label{3b}
We used the $^{13}$C$_2$ (second qubit) as the working medium
and  the $^{13}$C$_1$(first qubit) as the auxiliary qubit to
implement the isochoric heating stroke.
The first stroke of the QHE initializes the qubits 
to the desired
spin temperature, such that the working qubit is at a cold
spin temperature and the auxiliary qubit is at a hot spin
temperature.
The effective spin temperature($\beta$) of the initial
Gibbs state is related to the ground state($p_0$) and the excited
state($p_1$) populations as\cite{Assis_2019}:
\begin{equation}
\frac{1}{k_BT}=\beta=\dfrac{1}{h\nu_i}ln\left(\frac{p_0}{p_1}\right)
\end{equation}
where $h$ is Planck's constant, $k_B$ is Boltzmann's constant, $T$ is the spin
temperature and $\nu$ is the frequency of the energy gap.  By applying an RF
pulse of an appropriate angle of rotation (between 0 and $\pi$) to the
pseudopure state, the populations are redistributed between the qubit states
and coherence is created.  A pulsed field gradient(PFG) is employed to destroy
unwanted coherence and achieve the final desired thermal state.
Thus, each qubit is
equilibrated at different ``pseudo-spin'' temperatures. We
perform quantum state tomography to reconstruct the final
state and to verify that the system is initialized to the required
thermal state.
We initialized the working qubit to a cold spin
temperature of $k_BT_C$=11.94 peV and the auxiliary qubit to
spin temperatures of
$k_BT_H^1$=6.45 peV and $k_BT_H^2$=8.45 peV, respectively.
  
During expansion, the original Hamiltonian for the system is
given by:
\begin{equation}
\mathcal{H}^{{\rm exp}}_0(t)=\dfrac{\hbar}{2}  2\pi[ \nu_x
\sigma_x + \nu_z(t) \sigma_z]
\end{equation}
with
$\nu_z(t)=D_{{\rm exp}}\dfrac{t^2}{\tau^2}
\left(\dfrac{1}{2}-\dfrac{t}{3\tau}\right)$.
The RF field intensity is adjusted such that $\nu_x$=1000 Hz
and $\nu_z(t)$ changes from 0 Hz at t=0 to 2500 Hz at
t=$\tau$. The driving time $\tau$ is varied in this
interval from
200 to 2250 $\mu$~s. The H$_{CD}$ term to perform STA
during expansion is given by Eq.\ref{H_cd}, using
the $\nu_x$ and $\nu_z(t)$ given above. This driving time is
much shorter than the decoherence time (which is in the
order of seconds) and hence the driving
process can be considered to be a unitary evolution
U$_{\tau,0}$\cite{Batalhao_2015,Peterson_2019}. The
expansion stroke is implemented by applying a unitary
GRAPE pulse (with a fidelity $\ge$ 0.999)
of duration varying from 600 $\mu$~s to 6000 $\mu$~s.
 
The isochoric heating stroke is performed by emulating the
heat exchange between
the working system and the bath using the $^{13}C_1$ 
spin, which was initially prepared in a pseudothermal state
with inverse spin temperature $\beta_H$. The thermalization
process is achieved by applying a SWAP gate between the
working and the auxiliary qubits (depicted  in the blue
shaded portion of Fig.~\ref{ckt_diag}).
 
The compression stroke is realized by implementing the
following Hamiltonian:
\begin{equation}
\mathcal{H}^{{\rm comp}}_0(t)=\mathcal{H}^{{\rm exp}}_0(t)
\end{equation}
with
$\nu_z(t)=C_{{\rm comp}}+D_{{\rm comp}}
\dfrac{t^2}{\tau^2}\left(\dfrac{1}{2}-\dfrac{t}{3\tau}\right)$,
such that $\nu_z(t)$ changes from 2500 Hz at $t=0$ to 0 Hz at
$t=\tau$ and H$_{CD}$ changes accordingly. 
GRAPE pulses have been used to implement the compression stroke operations.
The total duration of the heat engine operation is less than 30
ms, which is well below the transverse relaxation time for both the nuclear
spins. We have chosen the spin temperatures 
and the energy gap frequencies such that
the heat engine working condition~\cite{Quan_2007}:
\begin{equation} 
\frac{T_H}{T_C} > \frac{\nu_f}{\nu_i} 
\end{equation} 
is satisfied, where T$_H$ and T$_C$ are the hot and cold bath temperatures
respectively, and
$\nu_i$ =1000 Hz and $\nu_f$ =2692.6 Hz are the minimum and maximum 
energy gap frequencies, respectively. 
The maximum achievable efficiency for the heat engine
operation is given as: 
\begin{equation} 
\eta=1-\frac{\nu_i}{\nu_f}.
\end{equation} 
Using these values of the maximum and minimum 
energy gap, we obtain the Otto limit efficiency of $\approx 0.629$. 
\section{Results \& discussion} 
\label{results}
The dynamics of the quantum Otto heat engine was studied using an LZ type
model, by changing the driving time ($\tau$) for the adiabatic process and the
hot source temperature, such that the cold bath temperature is kept fixed. The
working qubit $^{13}$C$_2$ is initialized to a cold spin temperature
$k_BT_C$=1.9 peV, while the auxiliary qubit $^{13}$C$_1$ was prepared at two
different hot spin temperatures $k_BT_{H1}$=6.45 peV and $k_BT_{H2}$=8.45 peV, 
to
carry out two different sets of experiments. The fidelity of the experimentally
prepared quantum states at each temperature was verified via quantum state
tomography.

The performance of a QHE can be analyzed through the behavior of its efficiency
and the output power delivered. Figure~\ref{effi_plots} depicts these figures
of merit, with the hot reservoir temperature being kept at 6.45 peV and 8.45
peV respectively, while keeping the cold reservoir temperature fixed.  The
engine efficiency at different driving times ($\tau$) is shown in
Figure~\ref{effi_plots}(a)-(b), where the maximum achievable efficiency i.e.
the Otto limit ($\approx$ 0.629) is shown by a solid black line. Dashed blue,
dotted-dashed green and dotted red curves represent the theoretically predicted
efficiencies for different driving times of the STA and non-adiabatic(NA) heat
engines respectively, while the experimentally obtained values with error bars
for each heat engine are plotted with the same color code. We observed that for
smaller driving times the NA heat engine does not produce any useful work as
the entropy production is very high. As we increase the driving time, the
irreversible entropy production is reduced, and after $\approx 1300 \mu$~s and
$\approx 900 \mu $~s for T$_{H1}$ and T$_{H2}$ respectively, useful work is
extracted and the efficiency is positive. If the driving time is increased
gradually, the efficiency reaches the Otto limit, which is due to the fact that
the process is getting closer to the ideal adiabatic process, since a decrease
in the entropy production leads to an increase in the efficiency.

The theoretically calculated efficiency of STA engine using both definitions
(Eqs.\ref{first_STA_def} and \ref{second_STA_def}) is shown in
Figs.~\ref{effi_plots} (a) and (b) (dot-dashed green and dashed blue curves).
We observed that if we use the first definition of efficiency, the efficiency
even at low driving times is positive, which implies a certain amount of work
can be extracted, leading to a positive output power.  However, for low driving
times there is no output power, hence we conclude that the first definition of
efficiency is not relevant for our experiments.  On the other hand, using the
second definition of efficiency ensures that for low driving times, both the
efficiency and power are zero and this definition hence correctly captures the
STA engine dynamics. 

We note here in passing that, if we use the second definition of efficiency
(Eq.\ref{first_STA_def}), then both the STA heat engine plots retain a finite
amount of efficiency for small driving times such as 10 $\mu$s. This indicates
the supremacy of the STA protocol over the normal NA driving protocol. The STA
efficiency saturates to the Otto limit when driving time is enhanced. The STA
heat engine efficiency is always greater than the NA heat engine efficiency,
until the point where they both merge and become equal to the Otto limit at
very high driving times. 

Figs.~\ref{effi_plots}(c) and (d) display the power for various driving times.
Black, blue and red curves are for theoretically calculated values of power for
adiabatic, STA, and NA heat engines, respectively. The experimental power
values corresponding to STA and NA heat engines are plotted with the designated
color along with error bars. We obtained a good match between experimental data
and theoretical calculations. The efficiency trend has been followed by power
as well, as we observe that the STA heat engine power is always greater than
the NA heat engine power. The vertical brown dashed-dotted line marks the
driving time at which the output power is maximum for STA heat engine. We
obtained maximum power at 1550 $\mu s$ and 1210 $\mu s$ for the STA heat engine
when the hot reservoir temperature is kept at T$_{H1}$ and T$_{H2}$,
respectively. We achieved maximum power at 2000 $\mu s$ and 1500 $\mu s$ for
the NA heat engine (not shown in the plots), when the hot reservoir temperature
is kept at T$_{H1}$ and T$_{H2}$, respectively.  Therefore, we obtain more power
for smaller driving times for the STA heat engine, which implies that the
overall performance of the STA heat engine is better than the NA heat engine
even after spending an extra amount of energy to maintain the adiabatic path.
We observed that at low driving times, the output power is zero for the STA
heat engine even though the efficiency is non-zero, which is due to the
definition of power for the STA heat engine. The STA cost is significantly high
at lower driving times, which is subtracted from the output work in
Eq.\ref{STA_power}. Hence, the power becomes zero at low driving times and it
becomes positive only after the STA cost is reduced to a certain level after a
particular driving time. 

The addition of an extra term in the original Hamiltonian to implement STA via
counter-adiabatic driving requires an additional amount of energy termed the
`STA cost' for the expansion and compression processes. This extra cost can be
calculated according to Eq.\ref{STA_cost}, and Fig.\ref{STA_cost_plots}(a)-(b)
shows the theoretically calculated STA cost during the expansion and
compression operations for two different sets of reservoir temperatures. We
observed that the STA cost for expansion is the same in both the plots as the
cold reservoir temperature and initial and final energy gaps are fixed. The STA
cost for compression is smaller for higher hot source temperatures (T$_{H2}$ $>$
T$_{H1}$) as compared to lower hot source temperatures. Hence, this reduction in
STA cost gets reflected in the efficiency, and the efficiency at small driving
times for higher hot source temperatures is more than the lower hot source
temperatures. One possible explanation of having higher STA cost for expansion
is the fact that the population difference between the two states is high
system state is initialized to cold spin temperatures, which increases the
probability of transitions between the states.  Hence, more amount of energy is
required to maintain the population difference throughout the adiabatic
expansion operation. When the system starts from a hot spin temperature, the
population difference between the higher and lower energy states is less than
that for cold spin temperature systems. Hence the transition probability
between the states is reduced. Therefore, less amount of energy is spent to
maintain the population difference during the compression operation. As T$_{H2}$
$>$ T$_{H1}$, the STA cost for compression is less when the hot reservoir is at
T$_{H2}$ as compared to T$_{H1}$. This could  be due to the fact that T$_{H2}$ 
has lower population difference as compared to T$_{H2}$. 
\section{Conclusions}
\label{concl}
We experimentally implemented an STA protocol based on the counter-adiabatic
driving technique for a LZ model quantum Otto heat engine on an NMR quantum
processor. We used different hot source spin temperatures and driving times to
investigate the performance of the quantum Otto heat engine by taking into
account the extra STA cost which has been spent to keep the system in
quasi-static equilibrium.  We compared two different definitions of efficiency
for the STA engine and argued the correctness of one definition over the other.
It can be concluded from figures of merit such as efficiency and power that the
performance of the STA heat engine is superior to the non-adiabatic heat
engine, as it can deliver more output power in less driving time. 
Our results are a step forward in the direction of improving the
performance and efficiency of physical realizations of quantum
heat engines.
\begin{acknowledgments}
All experiments were performed on a Bruker Avance-III 600 MHz FT-NMR
spectrometer at the NMR Research Facility at IISER Mohali. K.S.
acknowledges financial support from the Prime Minister's Research
Fellowship(PMRF) scheme of the Government of India. M.K. acknowledges
financial support from DST-INSPIRE fellowship. 
\end{acknowledgments}


%
\end{document}